\documentclass[aps,twocolumn,prl,showpacs,amsmath,amssymb,showkeys,10pt]{revtex4-1}
\usepackage{graphicx}
\usepackage{epstopdf}
\usepackage{braket}
\usepackage{hyperref}
\allowdisplaybreaks

\begin{document}

\title{Second-order coherence of fluorescence in multi-photon blockade}

\author{Th. K. Mavrogordatos}
\email[Email address(es): ]{themis.mavrogordatos@fysik.su.se;  th.mavrogordatos@gmail.com}
\affiliation{Department of Physics, Stockholm University, SE-106 91, Stockholm, Sweden}
\author{C. Lled\'{o}}
\email[Email address: ]{c.lledo.17@ucl.ac.uk}
\affiliation{Department of Physics and Astronomy, University College London, Gower Street, WC1E 6BT, London, United Kingdom}

\date{\today}

\begin{abstract}
 We calculate the second-order correlation function for the atomic fluorescence in the two-photon resonance operation of a driven dissipative Jaynes-Cummings oscillator. We employ a minimal four-level model comprising the driven two-photon transition alongside two intermediate states visited in the dissipative cascaded process, in the spirit of [S. S. Shamailov {\it et al.}, Opt. Commun. {\bf 283}, 766 (2010)]. We point to the difference between the output of a JC oscillator exhibiting two-photon blockade and the scattered field of ordinary resonance fluorescence, and discuss the quantum interference effect involving the intermediate states, which is also captured in the axially transmitted light. The spectrum and intensity correlation of atomic emission explicitly reflect the particulars of the cascaded model. 
\end{abstract}


\maketitle

The persistence of photon blockade is an exemplary property of the driven dissipative Jaynes-Cummings (JC) interaction substantiating the strong-coupling one-atom ``thermodynamic limit'' in which quantum fluctuations are in continual disagreement with the semiclassical system response~\cite{Carmichael2015}. Photon blockade refers to a term coined more than two decades ago~\cite{Imamoglu1997} to describe a situation where the absorption of $n$ photons blocks the absorption of any additional photon due to the presence of an appreciable excitation-dependent frequency detuning. The experiment of~\cite{Bishop2009}, realizing strong coupling between a microwave field and a superconducting qubit, focused on the nonlinear response of the vacuum Rabi resonance which splits into a doublet at high driving powers. A splitting of that kind, unique to the one-atom vacuum resonance, is a result of the Rabi oscillation induced by the drive between the ground and first excited dressed state of the JC interaction, forming an effective two-level system. A two-state truncation of the Hilbert space comprising the two aforementioned levels was employed in~\cite{Tian1992} to model the saturation of the vacuum Rabi resonance. This analysis of the two-state behavior predicts a Mollow triplet spectrum~\cite{Mollow1969} for the axially as well as for side-scattered light due to the so-called {\it dressing of the dressed states} (see Sec. 13.3.3. of~\cite{CarmichaelQO2}), together with the photon antibunching and squeezing encountered in the ordinary free-space resonance fluorescence (see Sec. 2.3.6 of~\cite{CarmichaelQO1}). The fluorescence spectrum and second-order correlation function for a drive tuned to the lower vacuum Rabi resonance have been measured at microwave frequencies in the circuit QED experiment of~\cite{Lang2011}. 

Extending now to the two-photon resonance, the authors of~\cite{Shamailov2010} used a four-state truncation to study the spectrum and intensity correlations of the axially transmitted spectrum demonstrating the transition between extreme photon bunching to antibunching as the transition saturates. On the experimental front, the recent implementation of two-photon blockade and the violation of the Cauchy-Schwarz inequality for the intensity correlation function of the axial light in a cavity QED setup~\cite{Hamsen2017} constitutes and important advancement in the investigation of multi-photon quantum nonlinear optics.

The second-order coherence of side-scattered light in the weak excitation limit of single-atom cavity QED presents substantial differences from free-space resonance fluorescence (see Sec. 16.1.5 of~\cite{CarmichaelQO2}). Here we take the topic of nonperturbative cavity-induced modifications to the atomic emission further by examining the intensity correlations of side scattering in the photon-blockade operation of the JC oscillator. The system density matrix $\rho$ for the driven dissipative JC interaction obeys the Lindblad master equation (ME)
\begin{equation}\label{eq:ME1}
\begin{aligned}
 \frac{d\rho}{dt}&=-i[\omega_0(\sigma_{+}\sigma_{-} + a^{\dagger}a)+g(a\sigma_{+}+a^{\dagger}\sigma_{-}),\rho]\\
 &-i[\varepsilon_d (a e^{i\omega_d t} + a^{\dagger}e^{-i\omega_d t}),\rho]\\
 &+\kappa (2 a \rho a^{\dagger} -a^{\dagger}a \rho - \rho a^{\dagger}a)\\
 &+\frac{\gamma}{2}(2\sigma_{-}\rho \sigma_{+} - \sigma_{+}\sigma_{-}\rho - \rho \sigma_{+}\sigma_{-}),
 \end{aligned}
\end{equation}
where $a$ and $a^{\dagger}$ are the annihilation and creation operators for the cavity photons, $\sigma_{+}$ and $\sigma_{-}$ are the raising and lowering operators for the two-level atom, $g$ is the dipole coupling strength, $2\kappa$ is the photon loss rate from the cavity, and $\gamma$ is the spontaneous emission rate for the atom to modes other than the privileged cavity mode which is coherently driven with amplitude $\varepsilon_d$ and frequency $\omega_d$. Throughout our analysis we take
\begin{equation}
g/\kappa \gg 1, \quad \varepsilon_d/g \ll 1 \quad \text{and} \quad  \gamma=2\kappa.
\end{equation}
This assumption places us in the low-excitation regime of very strong light-matter coupling where the energy levels of the excited JC couplets, $E_{n,\pm}=n\hbar \omega_0  \pm \sqrt{n} \hbar g$, are prominently split in relation to their width, admitting only perturbative correlations due to presence of the external drive. Multi-photon resonances follow the appearance of the vacuum Rabi peak in the spectrum as the drive amplitude is increased. The resonances first manifest themselves as sharp lines and subsequently saturate \textemdash{a} characteristic feature of a driven effective two-state system. In this work, we focus on the two-photon resonance, expressing the ME~\eqref{eq:ME1} in a truncated Hilbert space comprising the first four dressed JC states for low excitation.
\begin{figure}
\begin{center}
\includegraphics[width=0.3\textwidth]{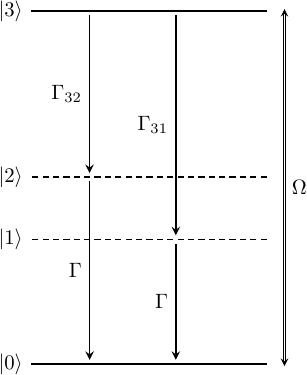}
\end{center}
\caption{{\it Schematic diagram of the four-level model.} The two-photon Rabi frequency $\Omega$ arises from a perturbative treatment due to the dressing of the dressed states by the drive. The dynamical evolution of the system described by this minimal model is governed by the effective ME~\eqref{eq:ME2}.}
\label{fig:levels}
\end{figure}
The JC dressed states between which transitions occur are depicted in Fig.~\ref{fig:levels} (see also Fig. 1 of~\cite{Hamsen2017}),
\begin{align}
 &\ket{0}\equiv\ket{0,-}, \\
 &\ket{1}\equiv \frac{1}{\sqrt{2}} (\ket{1,-} - \ket{0,+}), \\
 & \ket{2}\equiv \frac{1}{\sqrt{2}} (\ket{1,-} + \ket{0,+}), \\
 &  \ket{3}\equiv \frac{1}{\sqrt{2}} (\ket{2,-} - \ket{1,+}),
\end{align}
where $\ket{n, \pm}\equiv \ket{n}\otimes\ket{\pm}$, $\ket{n}$ is the Fock state of the cavity field, while $\ket{+}, \ket{-}$ are the upper and lower states of the two-level atom, respectively. Transitions to higher levels of the JC ladder are assumed far from resonance.   

We now transform ME~\eqref{eq:ME1} into a frame rotating by the driving frequency, $\rho \to U \rho U^{\dagger}$, with $U=\exp[i \hbar \omega_d t (a^{\dagger}a + \sigma_{+}\sigma_{-})]$. In this frame, the states $\ket{0}$ and $\ket{3}$ are degenerate and have zero energy when driving at the bare two-photon resonance, $\omega_d-\omega_0=-g/\sqrt{2}$, while the states $\ket{1}$ and $\ket{2}$ are retained despite being off-resonant since they are visited in the cascaded dissipative process. Keeping terms up to second order in the drive amplitude, we write the effective Hamiltonian in the rotating frame as (see Ch. 3 of~\cite{GottfriedQM})
\begin{equation}\label{eq:transformedH}
\begin{aligned}
 &\tilde{H}_{\rm eff}(E)=\mathcal{P} H_0 \mathcal{P} + \varepsilon_d \mathcal{P} (a+a^{\dagger}) \mathcal{P} \\
 &+ \varepsilon_d^2 \mathcal{P} (a+a^{\dagger}) \mathcal{Q}(E-\mathcal{Q}H_0 \mathcal{Q})^{-1} \mathcal{Q}(a+a^{\dagger}) \mathcal{P},
 \end{aligned}
\end{equation}
where $H_{0} \equiv \hbar (g/\sqrt{2})(a^{\dagger}a+\sigma_{+}\sigma_{-}) + \hbar g (a\sigma_{+} + a^{\dagger}\sigma_{-})$ and $E$ is the energy of a particular level, for which a perturbative correction is sought. In the Hamiltonian of Eq.~\eqref{eq:transformedH}, $\mathcal{P}$ projects onto the subspace with energy $E$, and $\mathcal{Q}=\mathbb{I}-\mathcal{P}$. The states $\ket{0}$ and $\ket{3}$ are then coupled via non-diagonal matrix elements as
\begin{equation}
\begin{aligned}
 \braket{3|\tilde{H}_{\rm eff}(E=0)|0}&=-\sum_{k=1,2}\frac{\varepsilon_d^2}{E_k}\braket{3|(a+a^{\dagger})|k}\braket{k|(a+a^{\dagger})|0}\\
 &=2\sqrt{2}\frac{\varepsilon_d^2}{g} \equiv \Omega. 
 \end{aligned}
\end{equation}
Likewise, we find perturbative corrections to the ground state energy, the intermediate couplet-state energies and the upper energy of the four-level scheme. Making the secular approximation in the limit of strong nonperturbative coupling ($g \gg \kappa, \gamma/2$) leads to the following effective ME when transforming back to the laboratory frame [see also Eq. (18) of~\cite{Shamailov2010}]
\begin{equation}\label{eq:ME2}
\begin{aligned}
  \frac{d\rho}{dt}&=\mathcal{L}\rho \equiv -i[\tilde{H}_{\rm eff},\rho]+\Gamma_{32} \mathcal{D}[|2\rangle \langle 3|](\rho) + \Gamma_{31} \mathcal{D}[|1\rangle \langle 3|](\rho)\\
  &+ \Gamma \mathcal{D}[|0\rangle \langle 1|](\rho) + \Gamma \mathcal{D}[|0\rangle \langle 2|](\rho),  
  \end{aligned}
\end{equation}
where
\begin{equation}
 \tilde{H}_{\rm eff}\equiv\sum_{k=0}^{3} \tilde{E}_{k} |k\rangle \langle k| + \hbar \Omega (e^{2i\omega_d t} |0\rangle \langle 3| + e^{-2i\omega_d t} |3\rangle \langle 0|),
\end{equation}
with the following shifted energies for the four states dressed by the drive,
\begin{subequations}\label{eq:dressedenergies}
 \begin{align}
 &\tilde{E}_0=E_0 + \hbar\delta_0(\varepsilon_d) = \hbar \sqrt{2} \varepsilon_d^2/g, \\
 & \tilde{E}_1=E_1 + \hbar\delta_1(\varepsilon_d) = \hbar \{\omega_0 -  g - [(20 + 19\sqrt{2})/7]\varepsilon_d^2/g\}, \\
 & \tilde{E}_2=E_2 + \hbar\delta_2(\varepsilon_d) = \hbar \{\omega_0 +  g +  [(20 - 19\sqrt{2})/7]\varepsilon_d^2/g\}, \\
 & \tilde{E}_3=E_3 + \hbar\delta_3(\varepsilon_d) =  \hbar (2\omega_0 - \sqrt{2}  g -  \sqrt{2}\, \varepsilon_d^2/g). 
\end{align}
\end{subequations}
In the effective ME of Eq.~\eqref{eq:ME2}, $\mathcal{D}[X](\rho)\equiv X\rho X^{\dagger}-(1/2)\{X^{\dagger}X, \rho\}$, and
\begin{align}
 &\Gamma_{31}\equiv\frac{\gamma}{4}+(\sqrt{2}+1)^2 \frac{\kappa}{2}=\frac{\gamma}{4}[1+(\sqrt{2}+1)^2], \\
 & \Gamma_{32}\equiv\frac{\gamma}{4}+(\sqrt{2}-1)^2 \frac{\kappa}{2}=\frac{\gamma}{4}[1+(\sqrt{2}-1)^2], \\
 & \Gamma\equiv\frac{\gamma}{2}+\kappa=\gamma,
\end{align}
are the transition rates between the dressed states comprising the four-level model depicted in Fig.~\ref{fig:levels}. In the dressed-state picture, truncated to the four states under consideration, the photon annihilation operator and the atomic lowering operator are
\begin{equation}
a \approx \frac{1}{\sqrt{2}} \ket{0}(\bra{1} + \bra{2}) + \frac{1}{2}[(\sqrt{2}+1)\ket{1} + (\sqrt{2}-1)\ket{2}]\bra{3}
\end{equation}
and 
\begin{equation}\label{eq:loweringop}
\sigma_{-} \approx -\frac{1}{\sqrt{2}} \ket{0}(\bra{1} - \bra{2}) - \frac{1}{2}(\ket{1} + \ket{2})\bra{3},
\end{equation}
respectively. The atomic-excitation operator reads
\begin{equation}\label{eq:Atss}
 \sigma_{+}\sigma_{-} \approx \frac{1}{2}\left(|1\rangle \langle 1|+|2\rangle \langle 2|-|2\rangle \langle 1| - |1\rangle \langle 2|+ |3\rangle \langle 3| \right).
\end{equation}
Since the density-matrix elements corresponding to the non-diagonal terms in Eq.~\eqref{eq:Atss} decay eventually to zero, the steady-state average atomic excitation is
\begin{equation}
\begin{aligned}
 \braket{ \sigma_{+}\sigma_{-}}_{\rm ss}&=\frac{1}{2}(p_1 + p_2 + p_3) =\frac{1}{2} \left(\frac{\Gamma_{31}}{\Gamma}+ \frac{\Gamma_{32}}{\Gamma}+ 1\right)p_3\\
 &=\frac{3\Omega^2}{2(\gamma^2 + 4\Omega^2)},
 \end{aligned}
\end{equation}
where we have used the expressions [$p_k \equiv (\rho_{kk})_{\rm ss}$, $k=0, 1, 2, 3$]
\begin{equation}
 p_1 = \frac{\Gamma_{31}}{\Gamma} p_3, \quad  p_2 = \frac{\Gamma_{32}}{\Gamma} p_3,
\end{equation}
linking the excited-state occupations as obtained from the exact solution of the effective ME~\eqref{eq:ME2}. In the strong-excitation limit, $\Omega^2 \gg \gamma^2$, we obtain $\braket{\sigma_{+}\sigma_{-}}_{\rm ss}=3/8$, which is lower than the maximum value possible for a driven two-level atom, $\braket{\sigma_{+}\sigma_{-}}_{\rm ss}=1/2$ (see Sec. 2.3.3 of~\cite{CarmichaelQO1}). In the weak-excitation limit, $\gamma^2 \gg \Omega^2$, the power radiated from the atomic fluorescence is proportional to $\Omega^2 \sim \varepsilon_d^4$, in contrast to the $\varepsilon_d^2$-dependence of free-space resonance fluorescence. 
\begin{figure}
\begin{center}
\includegraphics[width=0.47\textwidth]{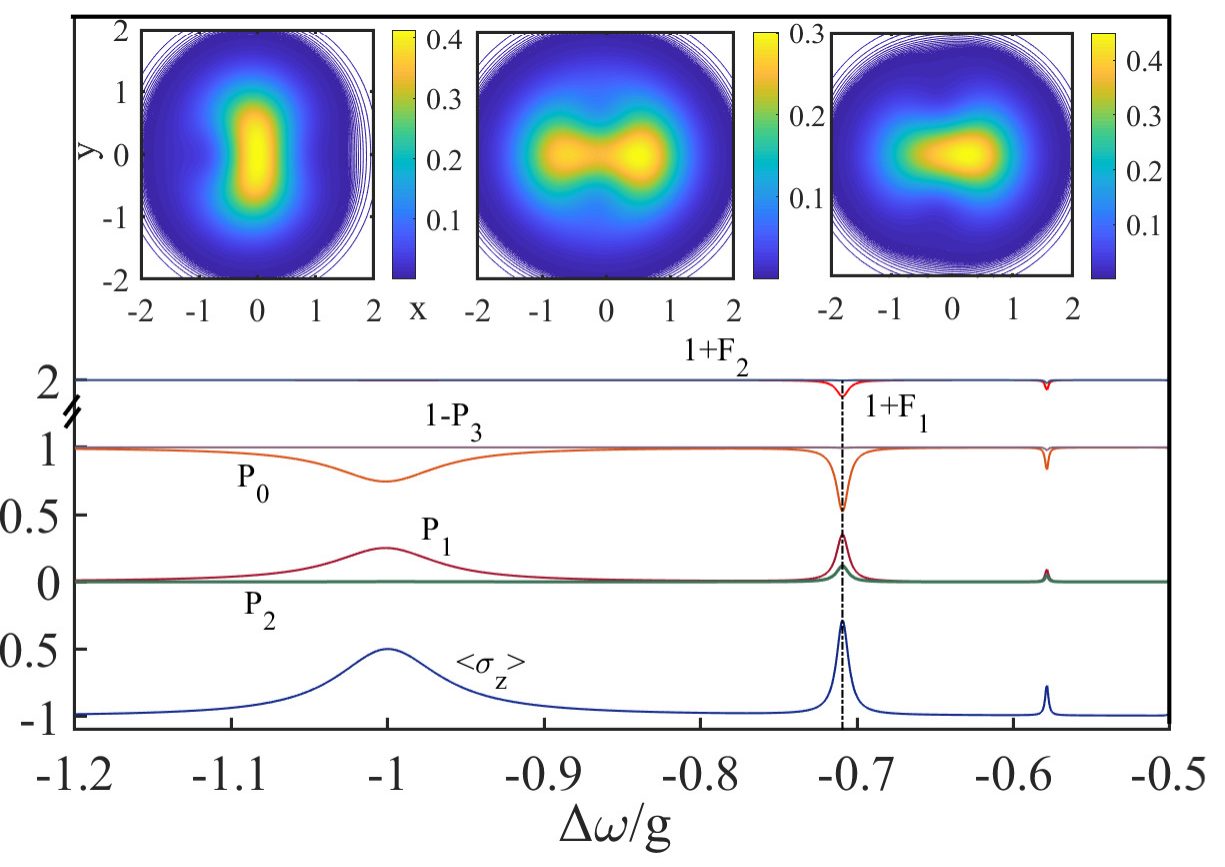}
\end{center}
\caption{{\it Multi-photon resonances for the coupled atom-cavity degrees of freedom.} The following steady-state quantities, computed from the ME~\eqref{eq:ME1}, are depicted going up the intersections of the corresponding curves with the dot-dashed line: average inversion of the two-level atom, $\braket{\sigma_z}_{\rm ss}$, probability of two, one and zero photon occupation ($P_2, P_1, P_0$), the complementary probability for the three-photon Fock state occupation ($1-P_3$) and, finally, the truncation fidelities $F_1, F_2$ as defined in Eq.~\eqref{eq:fid} (displaced by unity for visual clarity). The three insets on top depict the (steady-state) Wigner {\it quasi}probability distribution of the intracavity field, $W(x+iy)$, for $\Delta\omega/g=-1/1.40,-1/\sqrt{2},-1/1.42$ [on the left, center, and right, respectively]. The parameters used are: $g/\kappa=1000$, $\gamma=2\kappa$ and $\varepsilon_d/\kappa=40$.}
\label{fig:cavityandatom}
\end{figure}
Let us now look more closely at the phenomenon of photon blockade for the two coupled degrees of freedom comprising the JC oscillator. Further to the common picture of multi-photon resonances depicting the steady-state solution of Eq.~\eqref{eq:ME1} for the photon number, $\braket{a^{\dagger}a}_{\rm ss}$, we give an example of the operation at the two-photon resonance as displayed on the two-level atom in Fig.~\ref{fig:cavityandatom}; the steady-state solution of the ME~\eqref{eq:ME1} for the two-level inversion is plotted alongside various statistical quantities characterizing the intracavity field. The fidelity of an $m$-photon truncation is defined as~\cite{Miranowicz2013}
\begin{equation}\label{eq:fid}
 F_m(\rho_{\rm ss}) \equiv \sum_{n=0}^{m}P_n,
\end{equation}
where $P_n=\braket{n|\rho_{\rm ss}|n}$ are the $n$-photon Fock-state occupation probabilities in the steady state. If $F_m \approx 1$, then the contribution of Fock states with $(m+1)$ photons and above is negligible, which is a consequence of photon blockade.

In Fig.~\ref{fig:cavityandatom}, we plot the fidelities $F_1$ and $F_2$ on top of the steady-state atomic inversion [as obtained from the ME~\eqref{eq:ME1}] for a range of detuning in which the first three photon resonances are observed. At the detuning indicated by the dot-dashed line [$\Delta \omega/g=-1/\sqrt{2}-2\sqrt{2}(\varepsilon_d/g)^2\approx -0.71$] we observe that at most two cavity photons can be present in the JC oscillator, with $P_0=0.52$, $P_1=0.36$ and $P_2=0.12$, summing to the fidelity $F_2=1$. Hence, the single and two-photon Fock states block the generation of additional photons. We note that the vacuum Rabi resonance (centered around $\Delta\omega/g=-1$) has already saturated for the operating conditions considered. In the three insets of Fig.~\ref{fig:cavityandatom} we plot the Wigner function of the intracavity field as we trace through the second-photon resonance. In this process, we observe a rotation in the phase plane of the field and the occurrence of a bimodal distribution for $\Delta\omega/g=-1/\sqrt{2}$, a further evidence of the nonlinearity associated with photon blockade. The bimodal distribution remains in evidence as $\gamma/(2\kappa) \to 0$, where the two attractors are more clearly defined. In that limit of ``zero system size'' [$\gamma^2/(8g^2) \to 0$ \textemdash{see}~\cite{Carmichael2015} and Ch. 16 of~\cite{CarmichaelQO2}], we find that the left peak of the distribution (occurring at $\alpha \approx -0.68$) corresponds to a photon number matching the prediction of the neoclassical equations of motion~\cite{Carmichael2015}, along the upper branch of the input-output curve of bistability, rather than the two-photon Fock-state amplitude. Overall, however, the semiclassical picture provides a rather poor guide for understanding the response, since quantum fluctuations are inseparably intertwined with the system nonlinearity.

Following the attainment of steady state, a photon-emission event from the two-level atom creates the superposition [see second term of the atomic lowering operator in Eq.~\eqref{eq:loweringop}]
\begin{equation}\label{eq:superpos}
 \ket{\psi_{\rm super}}=\frac{1}{\sqrt{2}} (\ket{1}+\ket{2}),
\end{equation}
preparing the system in the following mixed state:
\begin{equation}
 \rho_{\rm cond}=\frac{\sigma_{-}\rho_{\rm ss}\sigma_{+}}{{\rm tr}(\sigma_{-}\rho_{\rm ss}\sigma_{+})}=\frac{2}{3}|0\rangle \langle 0| + \frac{1}{3}|\psi_{\rm super} \rangle \langle \psi_{\rm super}|.
\end{equation}
\begin{figure}
\begin{center}
\includegraphics[width=0.47\textwidth]{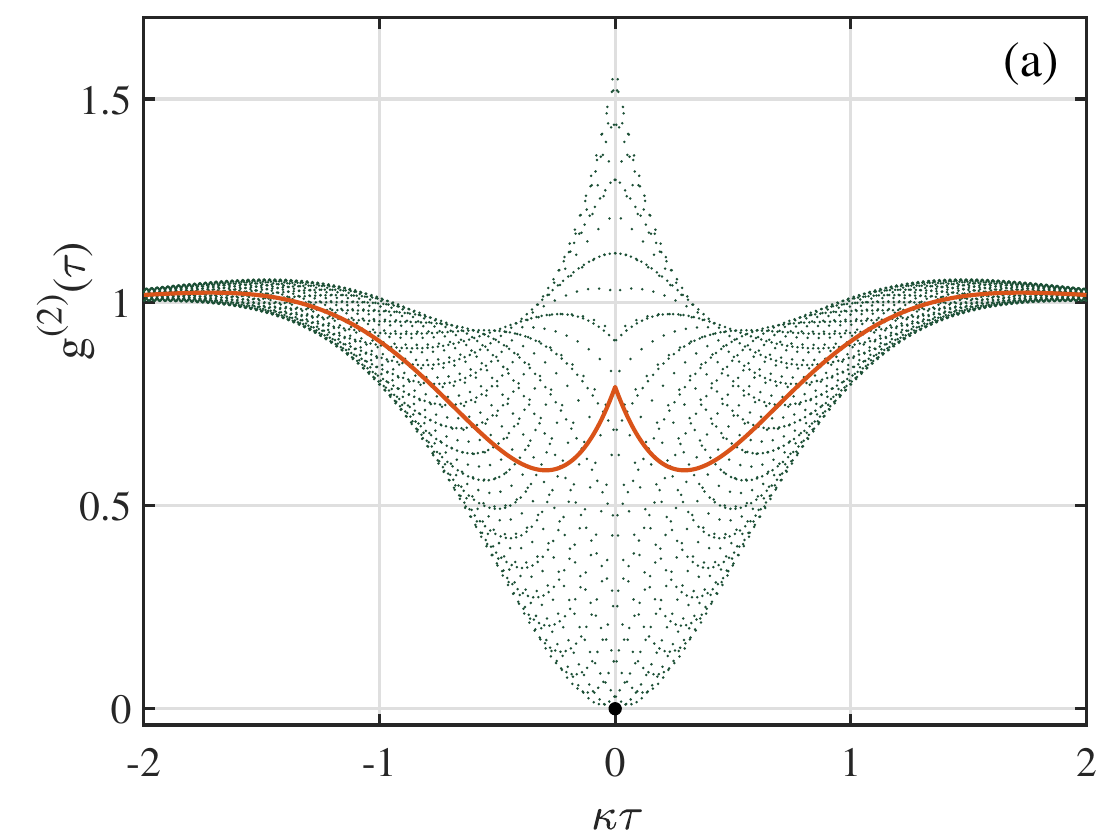}
\includegraphics[width=0.47\textwidth]{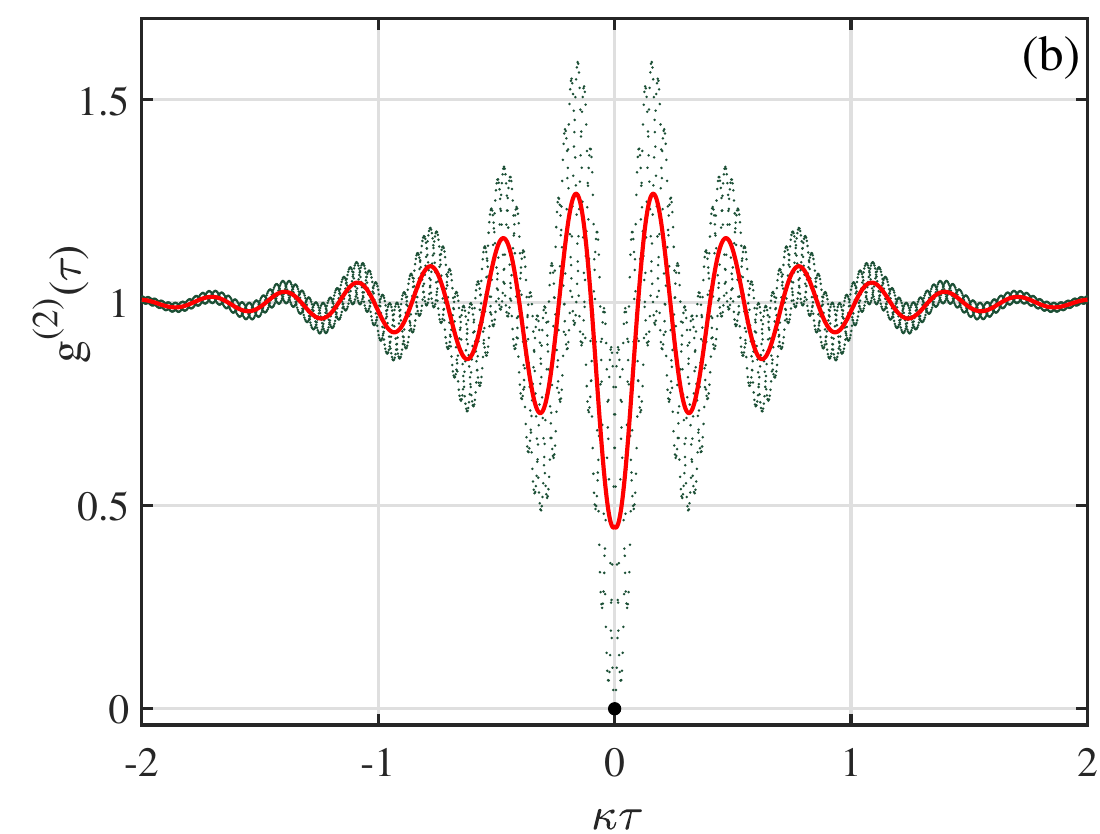}
\end{center}
\caption{{\it Second-order correlation function of side scattering for resonant excitation of the two-photon transition.} The function $g^{(2)}(\tau)$ of Eq.~\eqref{eq:g2final} is plotted for $g/\kappa=1000$, $\gamma=2\kappa$ and: {\bf (a)} $\varepsilon_d/\kappa=20$, {\bf (b)} $\varepsilon_d/\kappa=60$. We plot the full expression in green dots, while we use a solid orange line to signify an averaged-out quantum beat [i.e., without the contribution of Eq.~\eqref{eq:qb}]. The thick dot marks out the value $g^{(2)}(0)=0$.}
\label{fig:g2}
\end{figure}
\begin{figure*}
\begin{center}
\includegraphics[width=\textwidth]{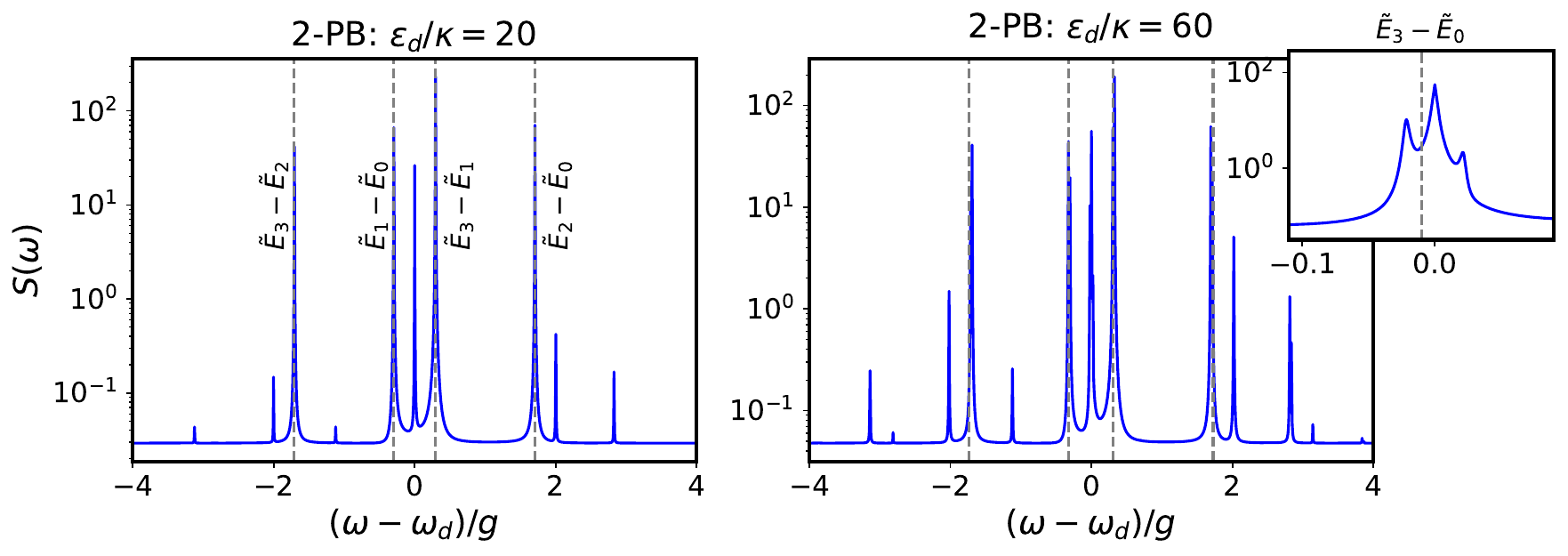}
\end{center}
\caption{{\it First-order coherence of fluorescence in two-photon blockade.} The fluorescence spectrum calculated from $S(\omega)=(1/\pi) {\rm Re}\{\int_{0}^{\infty} e^{i\omega \tau}\braket{\sigma_{+}(0)\sigma_{-}(\tau)}_{\rm ss}\,d\tau\}$ is plotted for $g/\kappa=1000$, $\gamma=2\kappa$ and: $\varepsilon_d/\kappa=20$ ({\bf left}) and $\varepsilon_d/\kappa=60$ ({\bf right}), as indicated on top of each frame. The dashed lines mark the dressed-state energy differences from Eqs.~\eqref{eq:dressedenergies}. The inset on the upper-right corner focuses on the Stark splitting of the driven two-photon transition.}
\label{fig:spfl}
\end{figure*}
The intensity correlation function is obtained by evolving the above mixed state forwards in time in accordance with the effective ME~\eqref{eq:ME2}, evaluating the mean atomic excitation in the time-evolved state, and normalizing by the steady-state excitation (denoted by the subscript ${\rm ss}$),
\begin{equation}\label{eq:g2A}
\begin{aligned}
 g^{(2)}(\tau)& \equiv \frac{\braket{\sigma_{+}(0)\sigma_{+}(\tau)\sigma_{-}(\tau)\sigma_{-}(0)}_{\rm ss}}{\braket{\sigma_{+}\sigma_{-}}_{\rm ss}^2}\\
 &=\frac{{\rm tr} \left\{[e^{\mathcal{L}\tau}\rho_{\rm cond}] \sigma_{+}\sigma_{-}\right\}}{\braket{\sigma_{+}\sigma_{-}}_{\rm ss}}\\
 &= \frac{1}{2}\frac{[\rho_{11}(\tau)+\rho_{22}(\tau)-\rho_{12}(\tau)-\rho_{21}(\tau)+\rho_{33}(\tau)]}{{\braket{\sigma_{+}\sigma_{-}}_{\rm ss}}},
 \end{aligned}
\end{equation}
with initial conditions set by $\rho(0)=\rho_{\rm cond}$. Two non-diagonal density-matrix elements feature in the expression of Eq.~\eqref{eq:g2A}, $\rho_{12}(\tau)=\rho_{21}^{*}(\tau)$, which can be determined independently of the rest in the secular approximation. They satisfy the equation
\begin{equation}
 \dot{\rho}_{12}\equiv \frac{d{\rho}_{12}}{d\tau}=-(i/\hbar)(\tilde{E}_1 - \tilde{E}_2) \rho_{12} - \Gamma\rho_{12},
\end{equation}
with initial condition $\rho_{12}(0)=1/6$. Hence, the quantum beat, arising from the superposition of states $\ket{1}$ and $\ket{2}$ following an atomic emission event, has the following contribution in the intensity correlation function:
\begin{equation}\label{eq:qb}
 g_{\rm qb}(\tau)=-\frac{\gamma^2 + 4\Omega^2}{9\Omega^2}e^{-\gamma \tau} \cos (\nu \tau),
\end{equation}
where $\nu \equiv 2g + \delta_2(\varepsilon_d) - \delta_1(\varepsilon_d)$ is the beat frequency, setting the smallest time scale in our problem. The remaining density-matrix elements in the expression of Eq.~\eqref{eq:g2A} defining the intensity correlation function obey the equations of motion
\begin{align}
 \dot{\rho}_{00}&=\gamma(\rho_{22}+\rho_{11})-i\Omega (\rho_{30}e^{2i\omega_d \tau}-\rho_{03}e^{-2i\omega_d \tau}), \label{eq:system1A} \\
    \dot{\rho}_{33}&=-2\gamma\rho_{33} -i \Omega (\rho_{03}e^{-2i\omega_d \tau}-\rho_{30}e^{2i\omega_d \tau}),\label{eq:system1B} \\
     \dot{\rho}_{03}&=-(i/\hbar)(\tilde{E}_0 - \tilde{E}_3) \rho_{03} - i\Omega(\rho_{33}-\rho_{00})e^{2i\omega_d \tau} -\gamma\rho_{03}\label{eq:system1C}
\end{align}
and
\begin{align}
   \dot{\rho}_{11}&=-\gamma \rho_{11} + \Gamma_{31} \rho_{33}, \label{eq:system2A}  \\
   \dot{\rho}_{22}&=-\gamma \rho_{22} + \Gamma_{32} \rho_{33}. \label{eq:system2B} 
\end{align}
Transforming the density-matrix elements as $\overline{\rho}_{ij}= e^{(i/\hbar)(\tilde{E}_i-\tilde{E}_j)\tau}\rho_{ij}$ and exciting the two-photon resonance with a drive frequency $\omega_d$, given by
\begin{equation}
 2\omega_d=(\tilde{E}_3-\tilde{E}_1)/\hbar=2\omega_0-\sqrt{2}g + \delta_3(\varepsilon_d)-\delta_0(\varepsilon_d),
\end{equation}
eliminates the fast-oscillating terms and reveals the energy scale defined by the system coupling rates. Based on the system of equations~\eqref{eq:system1A}--\eqref{eq:system1C}, we find that the vector $\boldsymbol{u} \equiv (D, D^{*}, \Sigma)^{\top}$, where $D \equiv \overline{\rho}_{03}$ and $\Sigma \equiv \overline{\rho}_{33}-\overline{\rho}_{00}=\rho_{33}-\rho_{00}$, obeys the equation
\begin{equation}\label{eq:ueom}
 \dot{\boldsymbol{u}}=\boldsymbol{M}\boldsymbol{u}+\boldsymbol{B},
\end{equation}
for $\boldsymbol{B}=(0,0,-\gamma)^{\top}$ and 
\begin{equation}
 \boldsymbol{M}=\begin{pmatrix}
 -\gamma & 0 & -i\Omega \\
 0 & -\gamma & +i\Omega \\
 -2i \Omega & +2i \Omega & -\gamma
\end{pmatrix}.
\end{equation}
The solution of Eq.~\eqref{eq:ueom} is
\begin{equation}
 \boldsymbol{u} = -\boldsymbol{M}^{-1}\boldsymbol{B} + {\rm exp}(\boldsymbol{M}t) \boldsymbol{u}(0)  + {\rm exp}(\boldsymbol{M}t) \boldsymbol{M}^{-1}\boldsymbol{B},
\end{equation}
with $\boldsymbol{u}(0)=(0,0,\Sigma(0))^{\top}$ and $\Sigma(0) \equiv p_3-p_0=-2/3$. From this last definition, we note that the presence of the intermediate levels $\ket{1}$ and $\ket{2}$ prevents the ``inversion'' of the effective two-level system associated with the two-photon resonance from reaching the value $-1$, as we would expect from a single-atom emission event. After solving for $\rho_{33}$ we substitute in Eqs.~\eqref{eq:system2A} and~\eqref{eq:system2B} to determine the solution of 
\begin{equation}
  \dot{\rho}_{11} +  \dot{\rho}_{22}=-\gamma(\rho_{11} + \rho_{22}) +2\gamma \rho_{33},
\end{equation}
with $\rho_{11}(0) + \rho_{22}(0)=1/3$. Collecting the various contributions to the intensity correlation and substituting to the expression of Eq.~\eqref{eq:g2A} we finally obtain
\begin{equation}\label{eq:g2final}
\begin{aligned}
  g^{(2)}(\tau)&=1 + e^{-\gamma|\tau|} [c_1 \cos(2\Omega \tau) + c_2 \sin(2\Omega |\tau|) \\
  &+ c_3 e^{-\gamma |\tau|} +c_4 \cos(\nu \tau)],
  \end{aligned}
\end{equation}
where
\begin{align}
 &c_1=  \frac{\gamma^2 - 4\Omega^2}{9\Omega^2}, \\
 &c_2=-\frac{5\gamma}{9\Omega}, \\
 &c_3=-\frac{1}{9}, \\
 &c_4=- \frac{\gamma^2 + 4\Omega^2}{9\Omega^2}.
\end{align}

We note that $g^{(2)}(0)=0$, as expected from single-atom emission since $\sigma_{+}^2=\sigma_{-}^2=0$ identically. The last term on the right-hand side of the correlation function in Eq.~\eqref{eq:g2final} is the quantum beat of Eq.~\eqref{eq:qb}. In the limit of weak excitation, $\gamma^2 \gg \Omega^2$, and for $\gamma \tau \gtrsim 1$,
\begin{equation}
g^{(2)}(\tau)=1 + \frac{\gamma^2}{9 \Omega^2}e^{-\gamma|\tau|} [1-\cos(\nu \tau)],
\end{equation}
which shows that large values of the intensity correlation are possible. In contrast to what happens for the two-level atom, the coefficient of the quantum-beat term in the intensity correlation function of the light emitted by the cavity (analogous to $c_4$) is positive, allowing for extreme values of forwards-scattered photon bunching, $g^{(2)}_{F}(0)\approx 4 \gamma^2/(25 \Omega^2) \gg 1$. In the limit of strong excitation, $\Omega^2 \gg \gamma^2$, we obtain
\begin{equation}
  g^{(2)}(\tau)=1 - \frac{1}{9}e^{-\gamma|\tau|} [4 \cos(2\Omega \tau) + e^{-\gamma |\tau|} +4 \cos(\nu \tau)].
\end{equation}

\begin{figure}
\begin{center}
\includegraphics[width=0.47\textwidth]{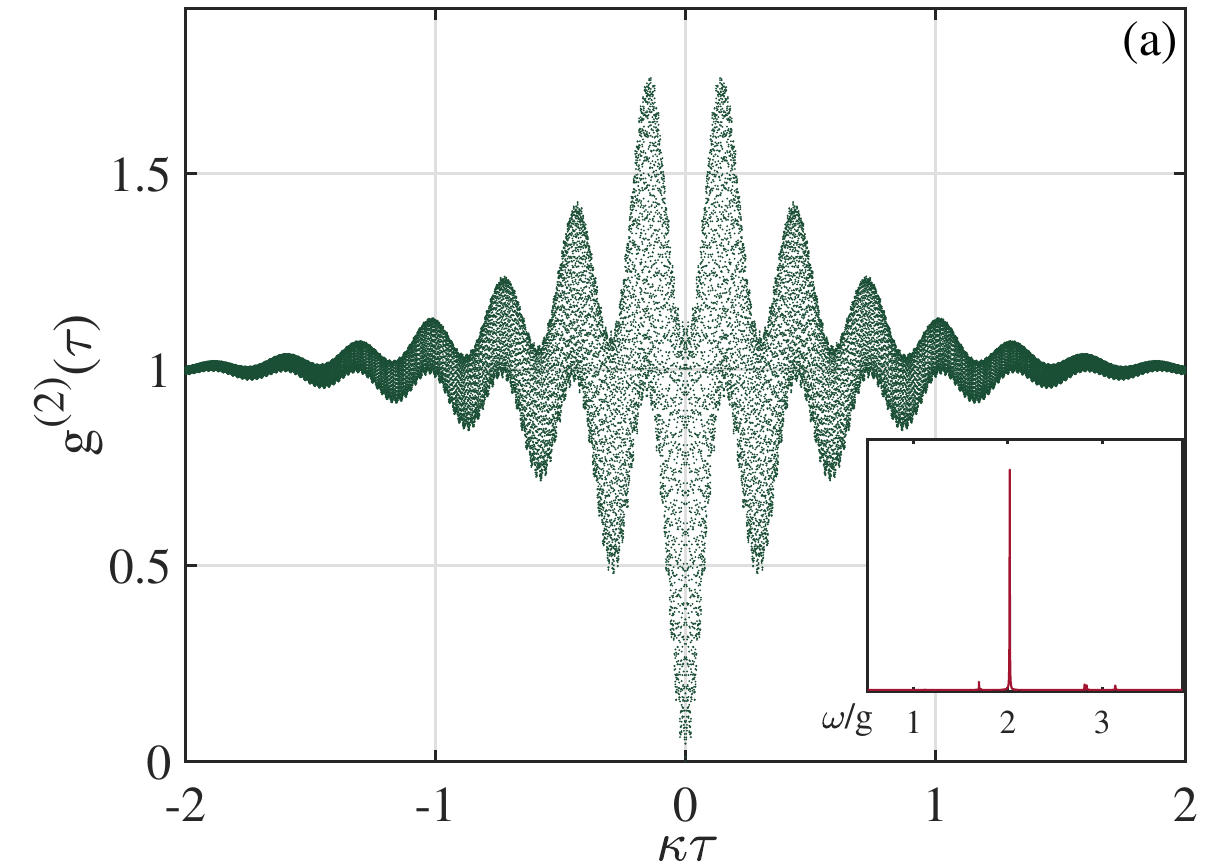}
\includegraphics[width=0.47\textwidth]{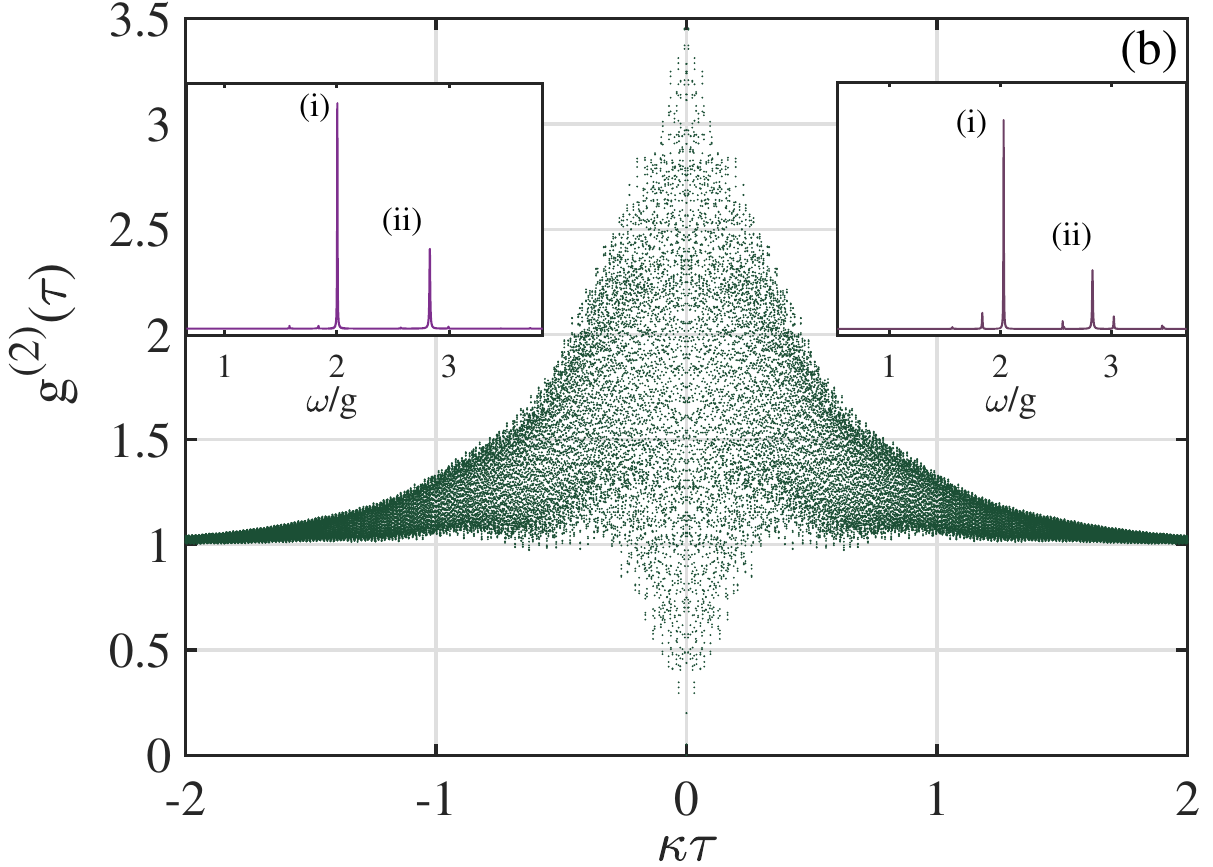}
\end{center}
\caption{{\it Numerically computed second-order correlation function of side scattering for the two-photon and three-photon resonances.} The function $g^{(2)}(\tau)$, extracted from the ME~\eqref{eq:ME1}, is plotted for $\varepsilon_d/\kappa=60$, $g/\kappa=1000$, $\gamma=2\kappa$ and: {\bf (a)} $\Delta \omega/g = -1/\sqrt{2}$, corresponding to Fig.~\ref{fig:g2}(b); {\bf (b)}  $\Delta \omega/g = -1/\sqrt{3}$. In frame (a), the inset depicts the magnitude of the Fourier transform of $g^{(2)}(\tau)$ against the dimensionless frequency $\omega/g$. In frame (b), the Fourier transform is plotted for $\varepsilon_d/\kappa=20$ and $\varepsilon_d/\kappa=60$ for the inset on the left and on the right, respectively. The peaks at $\omega \approx 2g$ and $\omega\approx 2\sqrt{2}g$ are denoted by (i) and (ii), respectively, while the peaks at frequencies $\sim \Omega$ are not shown as they are merged with the DC component.}
\label{fig:g2num}
\end{figure}

Evidence of the semiclassical Rabi splitting due to the saturation of the two-state of the effective two-state transition associated with photon blockade is given in Fig.~\ref{fig:g2} when progressing from frame (a) to frame (b) [compare also with Figs. 5(b-c) of~\cite{Shamailov2010} for the photon correlations of the axially-transmitted light]. In Fig.~\ref{fig:g2}(b), the quantum beat is superimposed on top of the semiclassical oscillations induced by the external drive, which now have a higher frequency on account of the increased saturation of the transition. In presenting Fig.~\ref{fig:g2}, we have separated the quantum beat of Eq.~\eqref{eq:qb} \textemdash{distinguishing} the one-photon from the multi-photon resonances \textemdash{from} the low-frequency terms comprising the intensity correlation function of atomic fluorescence. This distinction is largely artificial though. The term carrying the quantum beat must be kept to maintain the value $g^{(2)}(0)=0$, while for strong excitation it is superimposed with the same amplitude on the damped semisclassical Rabi oscillation of frequency $2\Omega \ll \nu$. When exciting the oscillator at the bare three-photon resonance, we find an additional dominant peak centered about $\omega \approx 2\sqrt{2}g$ in the Fourier transform of the intensity correlation function, whose amplitude remains below that of the peak at $\omega\approx 2g$ across the entire range of $\varepsilon_d/g \ll 1$ [see Fig.~\ref{fig:g2num}].

To obtain a direct evidence of the de-excitation pathways depicted in Fig.~\ref{fig:levels}, we have also numerically computed the incoherent fluorescence spectrum for the operating conditions of Fig.~\ref{fig:g2}, depicted in Fig.~\ref{fig:spfl}. For that calculation, we have employed the ME~\eqref{eq:ME1} with a Fock-state basis of 30 photon states (which is also the case for all the numerical results presented herein). For $\varepsilon_d/\kappa=20$, the distribution is shaped by five dominant peaks placed in the following sequence with increasing dimensionless frequency $\overline{\omega}\equiv (\omega-\omega_d)/g$: for $\overline{\omega}<0$, the first peak to appear corresponds to the energy difference $\tilde{E}_3-\tilde{E}_2$ and the second one to $\tilde{E}_1-\tilde{E}_0$. The $\tilde{E}_3-\tilde{E}_0$ peak is located at $\overline{\omega}\approx 0$, while for $\overline{\omega}>0$ the first peak closer to zero corresponds to the difference $\tilde{E}_3-\tilde{E}_1$ and the one further apart to $\tilde{E}_2-\tilde{E}_0$ [compare with Fig. 3 of~\cite{Shamailov2010}]. These transitions correspond to the de-excitation paths depicted in Fig.~\ref{fig:levels} and are described by the effective ME~\eqref{eq:ME2}. When going to $\varepsilon_d/\kappa=60$, the peak at $\overline{\omega}=0$ shifts and splits into a Stark triplet, while each of these four dominant peaks experiences a splitting ($\sim \Omega$) due to the hybridization of the states $\ket{0}$ and $\ket{3}$ induced by the drive [see the driving term in the effective ME~\eqref{eq:ME2}]. This is also observed in the incoherent intensity plots for the forwards scattered field [see Fig. 4 of~\cite{Shamailov2010}]. 

With driving-field amplitudes satisfying $\varepsilon_d/\kappa \gtrapprox 3$, we have numerically verified that for the zero-delay high-order correlation function of the forwards-scattered light, $g_F^{(n)}(0)$, we have $g_F^{(3)}(0)<g_F^{(2)}(0)$ when driving at the bare two-photon resonance ($\Delta\omega/g=-1/\sqrt{2}$); furthermore, $g_F^{(3)}(0)<1$ is satisfied for $\varepsilon_d/\kappa \gtrapprox 10$ and up to the range of applicability of the perturbative expansion. In particular, for the two-photon resonance  indicated by the dot-dashed line in Fig.~\ref{fig:cavityandatom} (where $F_2=1$), we find $g_F^{(3)}(0)<g_F^{(2)}(0)<1$; our results abide with the classification scheme for the occurrence of a two-photon blockade [$g_F^{(3)}(0)<1<g_F^{(2)}(0)$] given in Table II of~\cite{Miranowicz2019} for the window $10 \lessapprox \varepsilon_d/\kappa \lessapprox 26$, i.e. for weaker drive-field amplitudes than the one used for Fig.~\ref{fig:cavityandatom}.

Fig.~\ref{fig:g2num} depicts the numerically evaluated second-order correlation function for two values of the drive-cavity detuning selected to match the bare two-photon and three-photon resonances. A number of features are worthy of note. First, there is very good agreement between the analytical expression of Eq.~\eqref{eq:g2final} and the solution of the full master equation~\eqref{eq:ME1} for single-atom optical bistability \textemdash{where} no truncation in the energy levels has been made \textemdash{shown} in frame (a). Both functions evidence a high-frequency beat on top of a semiclassical oscillation; the latter has  frequency $2\Omega$, which is a factor of $\sim (\varepsilon_d/g)^2$ lower than the light-matter coupling strength. Second, the Fourier transform of both correlations has a dominant peak at $\omega \approx 2g$, corresponding to the energy difference $\tilde{E}_2- \tilde{E}_1$ which defines the frequency of the quantum beat. Here the amplitudes of the intermediate states are equal in the superposition following the detection of a side-scattered photon [see Eq.~\eqref{eq:superpos}], in contrast to what happens for forwards photon scattering where the beat has partial visibility. In frame (b), we observe the appearance of a subdominant peak at $\omega\approx 2\sqrt{2} g$, signifying the presence of a second quantum beat between the two states of the second excited-state couplet. The two peaks mentioned here are more clearly observed for relatively low values of $\varepsilon_d/\kappa$, setting the ground to anticipate a spectroscopic imprint of higher couplet states in the JC ladder, separated by $2\sqrt{n}g$, before photon blockade breaks down by means of a dissipative quantum phase transition~\cite{Carmichael2015}. The additional peaks are revealed as higher multi-photon resonances come into play. 

In this brief communication, we have focused on the low-excitation regime of very strong coupling in one-atom cavity QED where the presence of single photons creates a significant frequency shift in the spectrum, revealing the inherent nonlinearity of the JC interaction. In the operating region where dominant quantum fluctuations shape the response of the driven dissipative JC oscillator, we have derived an analytic expression for the second-order coherence function of the side-scattered light probing the saturation of the two-photon resonance, which has been explicitly modeled as a cascaded process. We have shown that the fluorescence spectrum together with the intensity correlation function uncover the cascaded process via which multi-photon blockade is organized. Spectroscopic evidence for the vacuum Rabi mode splitting with two photons is given in Fig. 4 of~\cite{Fink2008} in the very strong coupling regime of circuit QED, following the first observation of the two-photon resonance in cavity QED~\cite{Schuster2008}. Since atom-photon superposition states involving up to two photons are routinely within experimental reach in cavity and circuit QED, we expect that scattering records of the side-scattered light should be in position to verify the mixture of semiclassical and quantum features in the intensity correlations of the fluorescence emanating from a driven multi-photon resonance.

\begin{acknowledgments}
We are grateful to Prof. H. J. Carmichael for insightful discussions. Th. K. M. acknowledges the financial support of the  Swedish Research Council (VR) alongside the Knut and Alice Wallenberg foundation (KAW). C. L. acknowledges the financial support of the National Agency for Research and Development (ANID)/Scholarship Program/DOCTORADO BECAS CHILE/2017 - 72180352.
\end{acknowledgments}

\bibliography{bibliography}

\begin{center}
 *****
\end{center}

\end{document}